\documentstyle[aps,twocolumn,epsf]{revtex}

\newcommand{\vol}[1]{{\bf #1}}

\newcommand{\ttitle}[1]{{\it #1}}
\newcommand{\pretitle}[1]{{\em #1},}

\newcommand{\tselea}[1]{\label{#1}}
\newcommand{\tseleq}[1]{\label{#1}} 
\newcommand{\tbib}[1]{\bibitem{#1}} 
\newcommand{\tref}[1]{(\ref{#1})}
\newcommand{\tcite}[1]{\cite{#1}}

\newcommand{\tseepsffile}[1]{{\tt #1}}
\newcommand{\tseepsfxsize}[1]{{}}

\renewcommand{\tseepsffile}[1]{\epsffile{#1}}
\renewcommand{\tseepsfxsize}[1]{\epsfxsize=#1}

\newcommand{\tnote}[1]{}
\newcommand{\tpre}[1]{}

\newcommand{\href}[2]{#2}
\newcommand{\eprint}[1]{{#1}}
\newcommand{\iceprint}[1]{\href{http://euclid.tp.ph.ic.ac.uk/}{Imperial/TP/#1}}
\newcommand{\tseeprint}[2]{\href{http://euclid.tp.ph.ic.ac.uk/links/time/papers/#2.ps}{Imperial/TP/#1}}

\renewcommand{\href}[2]{{#2}{}}
\renewcommand{\eprint}[1]{\href{http://xxx.soton.ac.uk/abs/#1}{{\tt #1}}}

\renewcommand{\tpre}[1]{#1}

\newcommand{\half}{\frac{1}{2}}
\newcommand{\bea}{\begin{eqnarray}}
\newcommand{\eea}{\end{eqnarray}}
\newcommand{\beq}{\begin{equation}}
\newcommand{\eeq}{\end{equation}}

\newcommand{\nnel}{\nonumber \\ {}}

\newcommand{\taui}{\tau_i}
\newcommand{\tauz}{\tau}

\newcommand{\Bret}{B_{\rm ret}}
\newcommand{\Seff}{S_{\rm eff}}
\newcommand{\Sefft}{S_{\rm eff}^{(2)}}
\newcommand{\intdtk}{\int \frac{d^3\veck}{(2\pi)^3}  \;}
\newcommand{\intdtp}{\int \frac{d^3\vecp}{(2\pi)^3}  \;}

\newcommand{\intdfk}{\int \frac{d^4k}{(2\pi)^4}  \;}

\newcommand{\intdfx}{\int d^4x \;}

\newcommand{\intdtauz}{\int_{\taui}^{\taui-i\beta} \! \! \! \! \! \! \! \! d\tauz \;}

\newcommand{\veck}{\vec{k}}
\newcommand{\vecp}{\vec{p}}
\newcommand{\vecx}{\vec{x}}

\newcommand{\calB}{{\cal B}}
\newcommand{\calL}{{\cal L}}

\newcommand{\ebom}{e^{\beta \omega}}

\newcommand{\etdtom}{e^{-2i\tau\omega}}

\begin{document}

\draft

\preprint{\parbox{6cm}{{\tt Imperial/TP/97-98/69} 
\\ \eprint{hep-ph/9808383} } }

\title{The Unique Derivative Expansion for Thermal Effective
Actions\thanks{{\tt hep-ph/9808383}. Based on a talk given at
\href{http://theory.uwinnipeg.ca/tft98.html}{TFT98 
- the 5th Workshop on Thermal Field
Theory},
Regensburg, Germany, 10-14 August, 1998. Original overheads
available at
\href{http://euclid.tp.ph.ic.ac.uk/links/time/papers/reg/regtalk.ps}{http://euclid.tp.ph.ic.ac.uk/\symbol{126}time/papers/reg/regtalk.ps}
}}

\author{T.S.Evans\thanks{email: 
\href{mailto:T.Evans@ic.ac.uk}{\tt T.Evans@ic.ac.uk}}\thanks{{WWW: 
\href{http://euclid.tp.ph.ic.ac.uk/links/time}{\tt http://euclid.tp.ph.ic.ac.uk/\symbol{126}time} }}
}

\address{\href{http://euclid.tp.ph.ic.ac.uk/}{Theoretical Physics}, 
Blackett Laboratory, Imperial College,
Prince Consort Road, London SW7 2BZ,  U.K. }

\date{21st August 1998, revised 16th September 1998.}

\maketitle

\begin{abstract}
I show that there is a unique and well behaved 
derivative expansion of an effective action at finite
temperature. The result is true for all formalisms including the
popular Closed Time Path and Imaginary Time methods.
\end{abstract}

\pacs{PACS: 11.10.Wx, 11.10.Lm\tpre{. Preprint {\tt
Imperial/TP/97-98/69}}}


The effective action has been a powerful weapon in the arsenal of
quantum field theory for many years \tcite{Co,Fr,DGMP,MTW}.  In most
cases, effective actions are still too difficult to work with.  One
way of extracting physically useful results is to try a further
approximation, the derivative expansion.  In particular, for an
effective action of a scalar field,  the   lowest order term of such
an expansion is the effective potential or free energy, a foremost
tool in the study of the different phases of a theory.

However it  has been known for  some time \tcite{AT} that at any
non-zero temperature the derivative expansion is not well behaved in
that there is no one, unique answer
\tcite{AT,Fu,Fu2,TSEzm,TSEze,vW,TSEwpg,EEV,GH,We2,DH,AL,Das}.   This
then destroys the link between the Free energy and the potential
that controls the dynamical evolution of a scalar field such as one
has many pictures of phase transitions.  

Here I will describe how the existing calculations incorrectly use 
equilibrium Green functions to calculate such effective actions.
 When considering the behaviour of slowly varying fields in a
thermal bath, i.e.\ the system described by a derivative expansion
of an effective action, then one finds that one must use different
types of Green function.  This is then seen to give a unique
derivative expansion of the effective action.

The example I will consider has two fields, $\phi$ and $\eta$, whose
dynamics is described by the Lagrangian
\beq
\calL [\phi,\eta] = \half \eta \Delta^{-1}{\eta}  - \half g \phi
\eta^2 + \calL [\phi] 
\tseleq{Lscalar} 
\eeq 
One obtains an effective action for the $\phi$ field by integrating
out the $\eta$ field as follows
\bea 
Z & = & \int D\phi \; {D\eta} \;
\exp \{i \int d^4x \calL[\phi,{\eta}]  \} 
\\
& = & \int D\phi \;
\exp \{ i{\Seff[\phi]} + i \int d^4x \; \calL [\phi] \} 
\nonumber \\
{\Seff[\phi]} &=& {\frac{i}{2} {\rm Tr} \left\{ \ln \left[ 1 - g 
\phi(x)\Delta(x,y) \right] \right\} }
\tselea{Ldash} 
\eea
Here the integration is exact (a pure Gaussian integral in $\eta$)
which is not always the case.   Thus  the effective action for
$\phi$ is the pure $\phi$ classical part $\int d^4x \calL [\phi]$
plus $\Seff[\phi]$.  The $\Seff [\phi]$ contains {\em all} $\eta$
fluctuations - both quantum and statistical.  

Note that I have made no detailed specifications about the initial
configuration or the subsequent evolution of the $\phi$ fields.  All
I have specified is that it interacts with an $\eta$ field in a
particular way, and I will be assuming that the $\eta$ field is in
equilibrium.  The whole point of an effective action, in this case
for $\phi$, is that it describes the behaviour of just a {\em small}
part of a system but it  contains the effects of all fields.  We are 
then free to manipulate the effective action in a manner appropriate
to the physical problem.  For instance I can treat $\Seff [\phi]$ as
a description of a classical theory, or use it as the basis of a full,
non-equilibrium quantum path integral.  

In general  the aim is that $\Seff [\phi]$ will produce a simpler
description and allow for better (non-perturbative?) approximations.
 However the problem is that $\Seff [\phi]$ is still too complicated
- it is non-local in $\phi(x)$ as I will note below.  The solution
is to use further approximations, namely first to expand the $\ln$
and then to use a derivative expansion.  The latter will be the
focus of this work.  

The validity of such further approximations, and indeed the
usefulness of integrating out one field first rather than another,
must always be justified by the nature of the physical problem. For
instance if the $\phi$ field has a mass much less than the mass of
the $\eta$ field then it makes sense to first integrate out the slow
fluctuations of the $\eta$ field.  A standard example is the
Euler-Heisenberg effective action for electromagentic fields in  a
QED plasma where the $\phi$ field would become the photon field, and
the $\eta$ field the electron field
\tcite{DGMP,GH}.\tnote{Alternatively, one may want to be more
sophisticated and split the $\phi$ field into two pieces.  There
would be a fluctuating part which one will also try to integrate
out.  The other part might be equal to the expectation value of the
$\phi$ field or may be connected to a real physical classical
source.}   Given that the recipe I use to obtain an effective action
for the simple model \tref{Lscalar} is often appropriate in genuine
physical problems, I will use it on \tref{Lscalar} to illustrate
generic problems and my solution to them.

From \tref{Ldash}, the next step is to expand the $\ln$ and I find
\bea
\Seff[\phi] &=& \frac{i}{2} {\rm Tr} \left\{ \ln \left[ 1 - g 
\phi(x)\Delta(x,y) \right] \right\} 
\\
&=& \Seff^{(1)} + {\Sefft} + \ldots 
\\
&=& -\frac{ig}{2} {\rm Tr} \left\{ \phi(x) \Delta(x,y)  \right\}
\\
&&
-\frac{ig^2}{4} {\rm Tr} \left\{ \phi(x) \Delta(x,y) \phi(y)
\Delta(y,x) \right\} 
+\ldots
\tselea{dexp1}
\\
&=& 
\setlength{\unitlength}{0.05cm}
\raisebox{-1.0cm}{\begin{picture}(120,40)(0,0)
\put(0,0){\tseepsfxsize{2cm}\tseepsffile{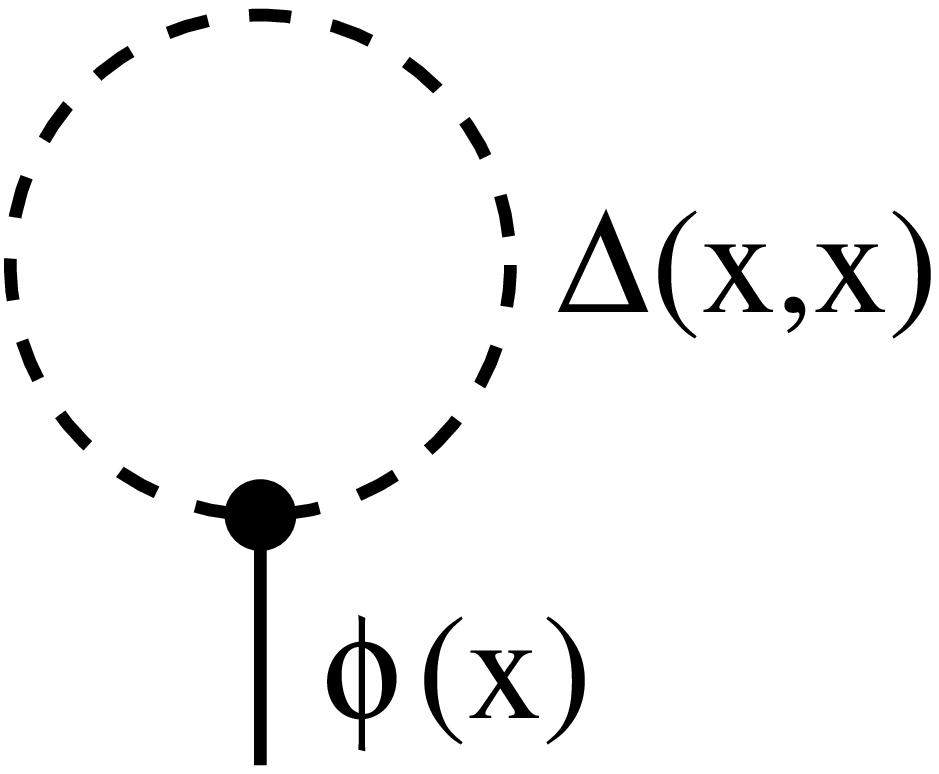}}
\thicklines
\put(40,10){+}
\put(50,0){\tseepsfxsize{3cm}\tseepsffile{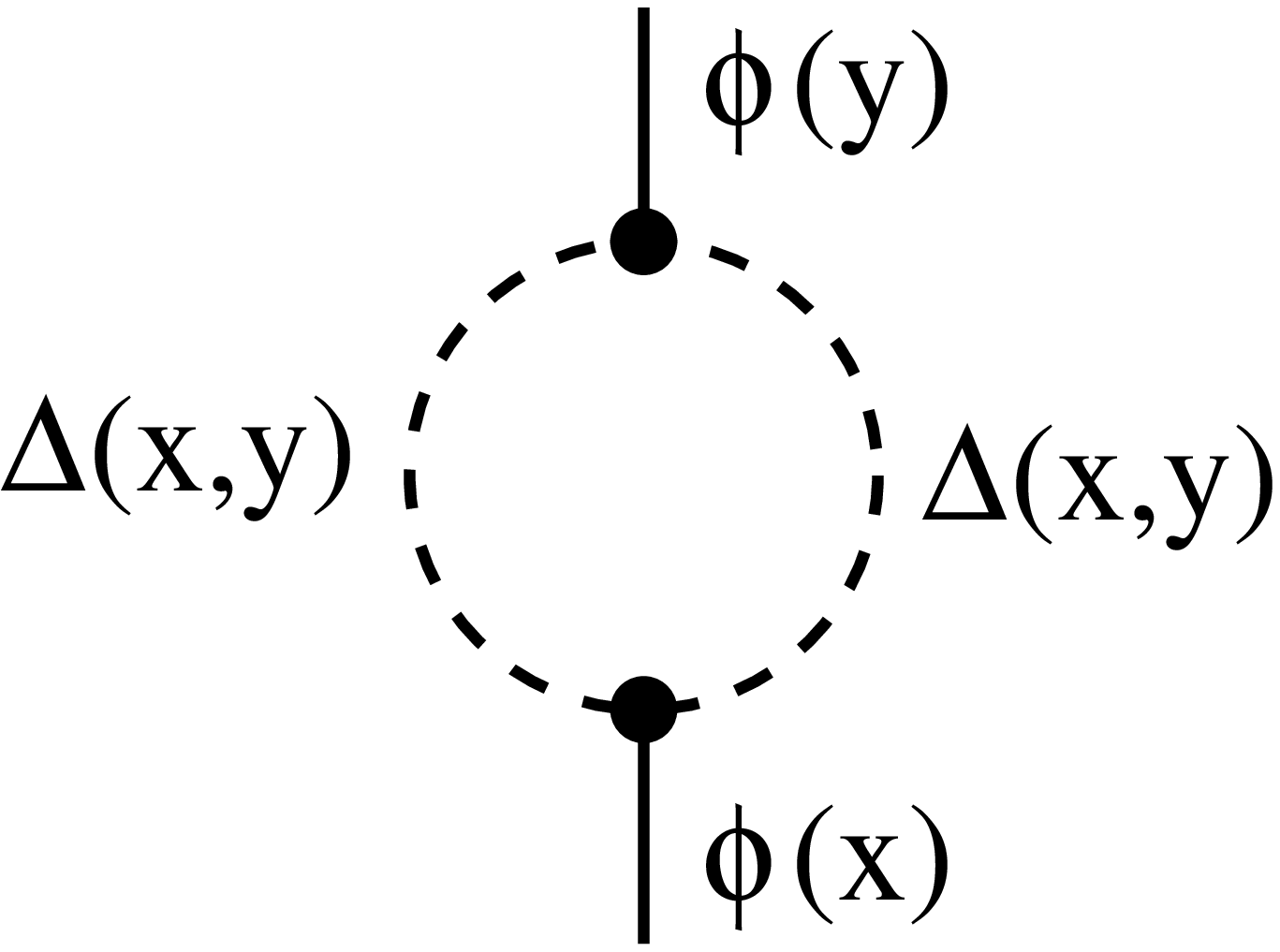}}
\put(108,10){+ ...}
\end{picture} }
\eea
where the dashed lines represent the $\eta$ field propagator, $\Delta$.
I will study the second term in this series, {$\Sefft$},
\beq
\Sefft =
 -\frac{ig^2}{4} \int d^4x \int d^4y  \left\{ \phi(x) \Delta(x,y) \phi(y)
\Delta(y,x) \right\}
\tseleq{Sefft}
\eeq
as this is the quadratic term in $\phi$ and so contains corrections
to the mass and kinetic terms of $\phi$.

{$\Sefft$} is still too complicated.  As classical actions
contain one space-time integral, I wish to keep one, say $\int
d^4x$, in {$\Sefft$} of \tref{Sefft}.  We need therefore to do the
second  $\int d^4y$ of \tref{Sefft}.  However, we see that we need
to know $\phi$ at both $x$ (considered fixed) and $y$ points and we
are integrating over all $y$ i.e.\ the effective action is
{non-local}.  This is too much information to expect.  We are more
likely to know a lot about the field at one point in time and/or
space.  It is therefore convenient in many problems to express the
field at any point in time or space, $y$, in terms of the field 
and its derivatives at one given point $x$.  Thus we use a Taylor
expansion to expand {$\phi(y)$} as an infinite series.
\bea
{\phi(y)} &=& {\phi(x)} + (y-x)^\mu \partial_\mu
{\phi(x)} + \ldots
\\
&=& \exp \{ i(y-x)^\mu p_\mu \} {\phi(x)}
\tselea{phiexp}
\eea
where $p_\mu = -i \partial_\mu \nonumber$. 
Then $\Sefft$ of \tref{Sefft} is given in terms
of an infinite number of $x$
terms, $\phi(x), \partial_\mu \phi (x)$ etc., the known $\eta$
propagator $\Delta(x-y)$,
and a series of simple $(y-x)_\mu$ polynomials.  This is the
derivative expansion.  Taken as a whole nothing is gained,
non-locality is replaced by the need to know an infinite number of
derivatives.  Progress is made by {\em truncating} this derivative
expansion so that we need specify only a finite number of local 
terms in a further approximation to our full effective action.

Under the above approximations the result is valid only in certain
physical situations.  The truncation of the $\ln$ expansion in 
\tref{dexp1} means that we are restricted to {weak
coupling} and/or {weak fields}.  The truncation of {derivative
expansion} means that we must consider fields varying  {\em slowly}
in time and space. 

Putting this all together gives the expression
\bea
\Sefft 
&=&
-\half
\intdfx  \phi(x) B(-i \partial_\mu ) \phi(x)
\tselea{dexp2}
\\
-i B(p) &=& \frac{(-ig)^2}{2}\intdfk i\Delta (k) i\Delta(k+p)
\tselea{bub}
\\
&=& \setlength{\unitlength}{0.075cm}
\raisebox{-1.0cm}{\begin{picture}(10,20)(0,0)
\put(0,0){\tseepsfxsize{0.75cm}\tseepsffile{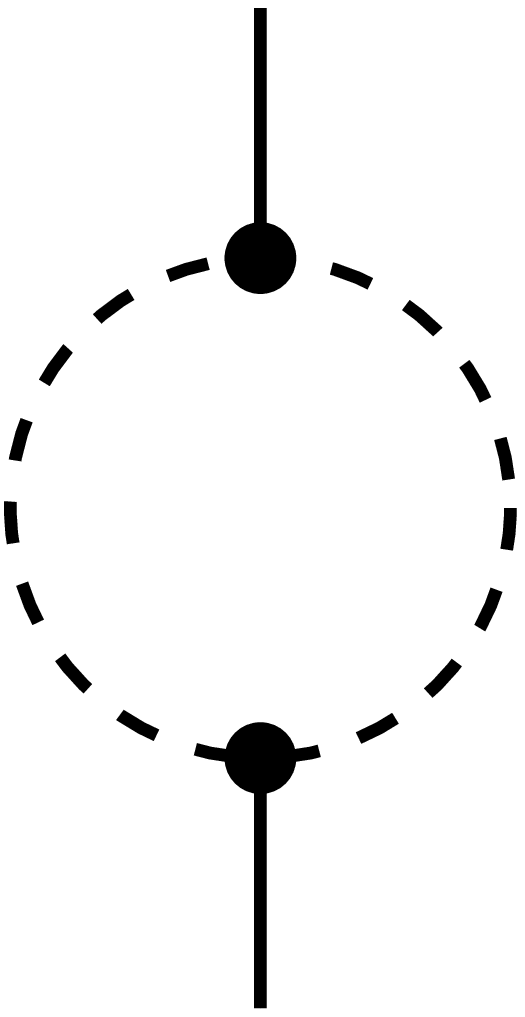}}
\end{picture}}
\eea
Comparing \tref{phiexp} with \tref{dexp2} shows that the derivative
expansion is an expansion in powers of  
$p^\mu = (E,\vecp)^\mu$. If, for exemplary purposes, I
just focus on time/energy $E \equiv -i \partial/\partial \tau$
then 
\bea
\Sefft &=&
\intdfx \left[  \right. B(p=0) \; \phi(x)^2 
\nnel
&&
+ \left(\frac{\partial B(p)}{\partial E} \right)_{p=0}
\phi(x) \frac{\partial}{\partial \tau}\phi(x)
\nnel
&& + \left.
\half  \left(\frac{\partial^2 B(p)}{\partial E^2} \right)_{p=0}
 \phi(x) \frac{\partial^2}{\partial \tau^2}\phi(x)
+ \ldots \right]
\eea
i.e.\ {\em the four-momentum expansion of
$B$ gives the derivative expansion of $\Sefft$.}

The analysis above is standard at zero temperature \tcite{Co,Fr,DGMP}.
Formally there
seems to be no problem at finite temperature, one can imagine that
the same Feynman diagrams are required but one just uses finite
temperature rules.  If one takes the retarded bubble diagram,
$B=\Bret$, then all thermal field formalisms give  
\bea
\lefteqn{\Bret (p_\mu=(E,\vecp)) = -\frac{g^2}{2}\intdtk
\sum_{s_0,s_1 =\pm 1}
\frac{s_0s_1}{4 \omega \Omega}
}
\nnel
&&
\frac{1}{E+s_0\omega + s_1\Omega}  \frac{(e^{\beta (s_0 \omega + s_1
\Omega)} - 1 ) }{ (e^{\beta s_0 \omega} -1)(e^{\beta s_1 \Omega}
-1)} 
\tselea{Bret} 
\eea 
where
\beq
\omega = [\veck^2+m^2]^{1/2}, \; \; \;
\Omega = [(\veck+\vecp)^2+m^2]^{1/2} .
\tseleq{omdef}
\eeq

However such equilibrium $B$'s do {\em not} have a unique
momentum expansion at any non-zero temperature!
This has been known for a long time, e.g.\ see Abrahams and Tsuneto
\tcite{AT}.  I learned this from Fujimoto \tcite{Fu2}.  A careful
calculation for the scalar retarded bubble gives
\tcite{TSEzm,TSEze,TSEbub}
\bea
\lefteqn{ \lim_{E,p \rightarrow 0} \Bret (E,\vecp) 
=
\mbox{(T=0 part)}
}
\nnel
&&- \; 
 \frac{g^2}{8 \pi^2} \int_m^{\infty} d\omega \; N(\omega)
\frac{ k}{\omega^2 + v^2\gamma^2 m^2}
\tselea{ebv}
\\
&\simeq& - \frac{g^2}{16 \pi} \frac{1}{1+\gamma} \frac{T}{m}
\tselea{ebvhit}
\eea
where
\bea
N(E) &=& [\exp(\beta E)-1]^{-1}
\tselea{bedef}
\\
v&=& \lim_{E,\vecp \rightarrow 0} \frac{|\vecp |}{E}, \; \; \;
\gamma = (1-v^2)^{-1/2}
\eea
Only in the case $v=\infty$ ($E \rightarrow 0$ before $\vec{p}
\rightarrow 0$) is the usual contribution to the effective potential
or Free energy is recovered.

The solution comes when we realise that we may {\em not} use just
the finite temperature Feynman rules in the zero temperature diagrams to
obtain a finite temperature result in a general calculation including
derivative expansions of effective actions.  Apart from anything
else, a Feynman diagram does {\em not} represent a unique function at
non-zero temperature there are many distinct types with
significantly different values (retarded, time-ordered, thermal Wightman,
\ldots) all of which all have their roles in
different physical problems \tcite{TSEnpt}.  One must always consider what the
physical problem is at finite temperature and let that tell you
which mathematical Green function can be used to extract the
relevant physics.

In the case of derivative expansions of effective actions, everyone
uses the {\em equilibrium} Green functions where {{\em all} fields}
are  assumed to be in equilibrium and periodic.\tnote{What about
RTF? Are fields periodic at all times?}  Usually retarded Green
functions are used \tcite{TSEwpg} though this is not always the case
\tcite{DH,Das}.  The question is {\em why are we using equilibrium
Green functions?}  A truncated derivative expansion of  a periodic
field is {\em not periodic}. Such approximations to field $\phi(y)$
are not equilibrium fields.
\begin{center}
\begin{tabular}{cccccl} 
{$\phi(\tau)$} &= & {$\phi(0)$} & +&
$\tau \; \; {\dot{\phi}(0)}$ & $ + \ldots$ \\
\mbox{periodic}      & & \mbox{constant,} & &\mbox{linear}  & \\
\mbox{in Im$(\tau)$} & &  \mbox{periodic} & &\mbox{{not periodic}} &  
\end{tabular}
\\ \mbox{ } \\
{\fbox{\parbox[t]{6.5cm}{\centering {So you MUST NOT assume that}
\\ {external legs represent equilibrium fields !}}}}
\end{center}
The Green functions relevant to derivative expansions of effective
actions are {{\em not}} generally equilibrium ones.  Of course! We
always wanted to look at how slowly varying fields (i.e.\ ones which
are clearly not in equilibrium) behave in an equilibrium {\em
background} - $\eta$ is in equilibrium, $\phi$ is not.

So let us repeat calculation but
\begin{itemize}
\item We will work with time not energy in the Feynman integrals. 
This means that we can do {\em all} path-ordered thermal formalisms easily.
\item The $\eta$ field is in equilibrium and I use the following form of the
propagator \tcite{lB} valid for all path-ordered thermal field theory
formalisms 
\bea
\Delta (\tau, \tau'; \veck) &=&
\frac{-i}{2\pi}\int_{-\infty}^{\infty} dE \; 
e^{-iE(\tau - \tau')}
\nnel
&& \left[ \theta_C(\tau ,\tau') + N(E) \right] \rho(E,\omega)
\eea
where $\theta_C$ is a contour theta function, $N(E)$ is
the Bose-Einstein distribution \tref{bedef}, and $\rho$
is the spectral function.  For the free scalar field $\eta$, $\rho$
is of the form
$(\pi / \omega).[ \delta(E-\omega) - \delta(E+\omega) ]$ with
$\omega$ given in \tref{omdef}.
\item The $\phi$ field is not assumed to be periodic or in equilibrium
so {$\exp \{ \beta E\} \neq 1$} etc.\ when $E$ is the energy or time
derivative associated with a $\phi$ field. I only use \tref{phiexp}
when doing the $\int dy$.
\end{itemize}
This gives
\bea
\lefteqn{ \Sefft =
-\frac{ig^2}{4} {\rm Tr} \left\{ \phi(x) \Delta(x,y) \phi(y)
\Delta(y,x) \right\} }
\\
&=& 
\setlength{\unitlength}{0.05cm}
\raisebox{-1.0cm}{\begin{picture}(40,40)(0,0)
\put(0,0){\tseepsfxsize{3cm}\tseepsffile{regbub.eps}}
\end{picture} }
\\
&=& -\frac{1}{2} \intdtauz \intdtp  
\phi(\tauz,\vecp)  {\calB ( E , p ; \tauz- \taui ) } \phi(\tau,
-\vecp) 
\nonumber
\eea
\bea
\lefteqn{ {\calB}  (E,\vecp; \tauz - \taui)  
}
\nnel
&&
= \frac{g^2}{2} \intdtk
\sum_{ s_0,s_1=\pm1}
\frac{ N(s_0\omega) N(s_1\Omega)}{4 (s_0\omega)(s_1 \Omega)} 
({F}+{G})
\tselea{Bea}
\\
& \neq & B_{ret}
\nonumber
\eea
\bea
{F} &=& \frac{(e^{\beta {A}} - 1)}{{A}} ,
\\
{G} &=&  \frac{(e^{-i(\tauz-\taui) {A}} -1 )}{{A}}
. \frac{e^{\beta (s_0 \omega +s_1 \Omega)} }{N(E)} 
\\
{A}&=&{E +s_0 \omega +s_1 \Omega}
\eea
where $E = -i\frac{\partial}{\partial \tau}$ and acts only on the
right hand side $\phi$ field.   The $\taui$ is the start of the path
in complex time used to define path-ordered thermal field formalisms
\tcite{lB}.  It is the time at which the density matrix is given in
closed-time path methods for instance.  
Since the derivative expansion requires an expansion of $\calB$
about $E,\vecp \rightarrow 0$ then we are interested in the
$\Omega \rightarrow \omega$ limit.  When $s_0=-s_1$ in
the sum in \tref{Bea} then $A \rightarrow 0$ and in this limit 
the denominators of $F$ and $G$ are zero.  These $s_0=-s_1$,
$\omega-\Omega$ terms are produced by Landau damping processes.  

This new result based on $\calB$ has several consequences:-\tnote{ Some
other things to think about.  First the $T=0$ limit.  This looks bad but
perhaps if I integrate by parts and/or look at the equations of motion,
there will be no problem.  Must do this.  Second, must mention the cuts
at $2m$ etc.  Must say that even though I have not apparently done the
expansion, the expression I have given is merely a memonic for such an
expansion.  Need to give an simple example to compare with e.g. expand
$(z-z_0)^{-1/2}$ and perhaps integrate this between $\pm 1$.}
\begin{enumerate}
\item $\calB$ gives {\em unique} derivative series.  When we
take $E,p \rightarrow 0$ then the 
{Landau damping terms} have $A \rightarrow 0$. However
both {F} and {G} are well behaved in this limit because of the form of
the numerators.
\item In contrast, the equilibrium $\Bret$ \tref{Bret} is equivalent to setting
$\exp \{\beta E \} =1 $ in $\calB$ \tref{Bea}\footnote{One must also
manipulate the pure $\omega$ dependent half of the integral in $\calB$, switching variables to
$\veck+\vecp$ and integrating over this to get the form
\tref{Bret}.},
so that in $\Bret$
$\exp \{{A} \} =\exp \{ \pm \omega \pm
\Omega\}$.  Thus when we look at the Landau damping terms in $\Bret$
where
$A\rightarrow 0$  we find that 
{$G$} $\rightarrow 0$, but {{$F$} $\sim$ {$A^{-1}$}}.  This
divergence in the integrand (there is still an integration over loop
momenta to be done) leads to the poor behaviour of the four-momentum
expansion of $\Bret$.
\item The only point where the correct $\calB$ and old retarded equilibrium $\Bret$
agree is in the {\em static} limit -
$E=0$.  This is the only case where the derivative expansion of the
field is in terms of a periodic field as a truely static
i.e.\ time-independent, is trivially periodic.  Another way of
viewing this result is to say that  this justifies the often ad-hoc
rule used in some early effective potential/free energy
calculations where the zero energy limit was taken first
\tcite{TSEwpg}. 
\end{enumerate}

We can now write out the derivative expansion.  Limiting myself to
first order in derivatives then I find that for {\em any}
path-ordered thermal field formalism the effective quadratic part of
the effective Lagrangian $\calL$ is
\bea
{\calL_{\rm eff}[\phi]} &=& (M^2+{\delta M^2}) \phi^2(\tau,\vecx)
\nnel
\; \; \;
&& -i\beta  \left( \half \delta M^2 + {\mu(\tau-\tau_i)} \right)
\phi(\tau,\vecx) \frac{\partial }{\partial \tau} \phi(\tau,\vecx)
\nnel
&&
\hspace{2cm} + \ldots
\\
{\delta M^2}  & = &
 - \frac{g^2}{8\pi^2} \int_0^\infty dk \; \frac{N(\omega)}{\omega}
\simeq
- \frac{g^2}{16\pi} \frac{T}{M} 
\nonumber \\
{\mu (\tau)} & {=} &
{ \frac{g^2}{32\pi^2}
\int_0^\infty dk \; N^2(\omega) \frac{k^2}{\omega^3}
\left[ -2i  \tau \omega \ebom 
\right. }
\nnel
\; \; \;
&& { \left. + 
\half (\etdtom -1 )(1+ e^{2i(\tau -i\beta) \omega})  
\right] }
\nonumber
\eea
I have assumed that the classical $\calL[\phi]$ of \tref{Lscalar}
gives an $M^2
\phi^2$ contribution to this order.  The key point is that this is
unique and independent of the order in which we take the zero
energy,momentum/time,space derivative limit.  It includes
contributions from Landau Damping terms which cause problems in
the zero momentum limits of equilibrium Green functions, yet there
is no problem when using $\calB$.  In fact
equilibrium retarded Green functions such as $\Bret$ describe the
response to a system to a delta function impulse, as this is what the
usual linear response analysis tells us \tcite{lB}.  Thus such Green
functions are 
relevant only for {\em sudden} changes.  The derivative expansion of
an effective action and $\calB$
are appropriate for slowly varying fields, a very
different type of physical problem.\tnote{Note 
that there is a linear $\beta$ term in the linear time derivative term. 
This appears to blow up badly as $T \rightarrow 0$ but in fact it does not 
contribute to the equations of motion, i.e. it can be removed by  
integration by parts.  I suspect this corresponds to the possibility of
doing the derivative expansion in time about various points in time, not just
the time of one of the fields.}

This analysis of the derivative expansion of an effective action 
for scalar fields has used a general path ordered approach to
thermal field theory using the two scalar field model of
\tref{Lscalar}. The same model has also been considered in detail
for the pure imaginary-time formalism \tcite{TSEea} and in the
closed-time path method \tcite{AG,Atalk}.  The same result emerges
in both cases as it must.  However the result is viewed very
differently in the two distinct approaches.  Other more sophisticated
models also have been considered - scalars or an abelian gauge field
coupled to fermions in a heat bath, models with chemical potentials,
 and in effective actions for weakly coupled superconductors
\tcite{ESea} - with similar success.

I would like to thank the organisers of TFT98 for creating such an
exciting meeting where this talk was given.  I would like to thank
my collaborators \tcite{AG,Atalk,ESea}
on various extensions of this work and Ray Rivers for continuing
discussions.  Finally thanks to G.Aarts, I.Lawrie and C.van Weert 
for their perceptive comments on my talk at TFT98.


\end{document}